\documentclass{icrc2009}

\usepackage{graphicx}   
\usepackage[caption=false]{caption}    
\usepackage[font=footnotesize]{subfig} 
\usepackage{fixltx2e}
\usepackage{url}

\newcommand{\shorttitle}[1]%
{\markboth{Proceedings of the 31\MakeLowercase{$^{st}$} ICRC, {\L}\'{o}d\'{z} 2009}{#1} }
\newcommand{\etal}{\MakeLowercase{\textit{et al. }}} 

\begin{document}
\title{The connection between optical and VHE $\gamma$-ray high states in blazar jets}

\author{\IEEEauthorblockN{Elina J. Lindfors\IEEEauthorrefmark{1},
			  Riho Reinthal\IEEEauthorrefmark{1},
                          Daniel Mazin\IEEEauthorrefmark{2},
                          Kari Nilsson\IEEEauthorrefmark{1},
                          Leo Takalo\IEEEauthorrefmark{1},
                          Aimo Sillanp\"a\"a\IEEEauthorrefmark{1} and\\
                          Andrei Berduygin \IEEEauthorrefmark{1}
                          on behalf of the MAGIC collaroration}
                            \\
\IEEEauthorblockA{\IEEEauthorrefmark{1}Tuorla Observatory, Department of Physics and Astronomy, University of Turku, 21500 Piikki\"o, Finland}
\IEEEauthorblockA{\IEEEauthorrefmark{2}Institut de Física d'Altes Energies (IFAE), E-08193 Bellaterra (Barcelona), Spain}}

\shorttitle{Lindfors \etal Optical to VHE $\gamma$-ray connection}
\maketitle

\begin{abstract}
  MAGIC has been performing optically triggered Target of Opportunity
  observations of flaring blazars since the beginning of its
  scientific operations. The alerts of flaring blazars originate from
  Tuorla blazar monitoring program, which started the optical
  monitoring of the candidate TeV blazars in 2002 and has now
  collected up to six years of data on 30 blazars. These Target of
  Opportunity observations have resulted in the discovery of three new
  VHE $\gamma$-ray emitting blazars (Mrk~180, 1ES~1011+496 and S5~0716+714) and in addition the discoveries of BL~Lac and 3C~279 were
  made during a high optical state. In this talk we present a detailed
  analysis of the optical light curves which are then compared to MAGIC
  observations of the same sources. We aim to answer the question "Is
  there a connection between optical and VHE gamma-ray high states in
  blazars or have we just been lucky?"
 
  \end{abstract}

 
\section{Introduction}
Blazars are the most extreme type of Active Galactic Nuclei (AGN). In
these objects the dominant radiation component originates in a
relativistic jet pointing nearly towards the observer. Blazars show a
variable flux in all wave bands from radio to very high energy (VHE,
here defined as E$>$ 100 GeV) $\gamma$-rays. The relationship between
the variability at different wave bands appears to be rather complicated.
It is generally assumed that the energy spectrum is dominated at lower
energies (radio-UV and even up to X-rays in some sources) by
synchrotron radiation from relativistic electrons spiraling in the
magnetic field of the jet, and at higher energies (X-ray to VHE
$\gamma$-rays) by inverse Compton scattering. Seed photons for inverse
Compton scattering can be local or co-spatial synchrotron photons
(synchrotron self-Compton, SSC), or external photons (EC) originating
from the accretion disk \cite{DS}, emission-line clouds \cite{Sikora},
or the molecular torus \cite{Blazejowski}. In the simplest case both 
SSC and EC mechanisms predict a connection between optical and
$\gamma$-ray flares. Time-lags between optical and $\gamma$-ray flares
may be present due to several mechanisms (see
e.g. \cite{sokolov04,chatterjee08} for detailed discussion). The
time-lags and their relative order (an optical flare leading a flare in $\gamma$-rays or vice versa) are impossible to predict. The
connection between optical and $\gamma$-ray flares should also depend
on the synchrotron peak frequency of the source. If
the peak position is at a much higher frequency (in the X-ray domain) the
variability in the optical band is in some cases very
small and the synchrotron flares are best visible in the X-ray
band. This seems to be the case for the best studied TeV blazars
Mrk~421 \cite{lichti08} and Mrk~501 \cite{pian}. 
In principle, hadronic models can also explain correlated variability and cannot be excluded
as the emission mechanism of $\gamma$-rays.

The MAGIC collaboration is performing Target of Opportunity
observations of sources in a high flux state in the optical, and/or X-ray
band. Alerts of optical high states are sent by the Tuorla Blazar
Monitoring Program (\cite{takalo}, 
). These triggered MAGIC observations and 
resulted in discovery of VHE $\gamma$-rays from Mrk~180 \cite{mrk180}, 
1ES~1011+496 \cite{1011}, 
and S5~0716+714 \cite{0716}. 
In this paper, we present analysis of the optical light curves and compare them to the simultaneous MAGIC observations. We discus the implications of the results to the optical- VHE $\gamma$-ray connection.
\\

\begin{figure*}[!ht]
\includegraphics[angle=270, width=0.95\textwidth]{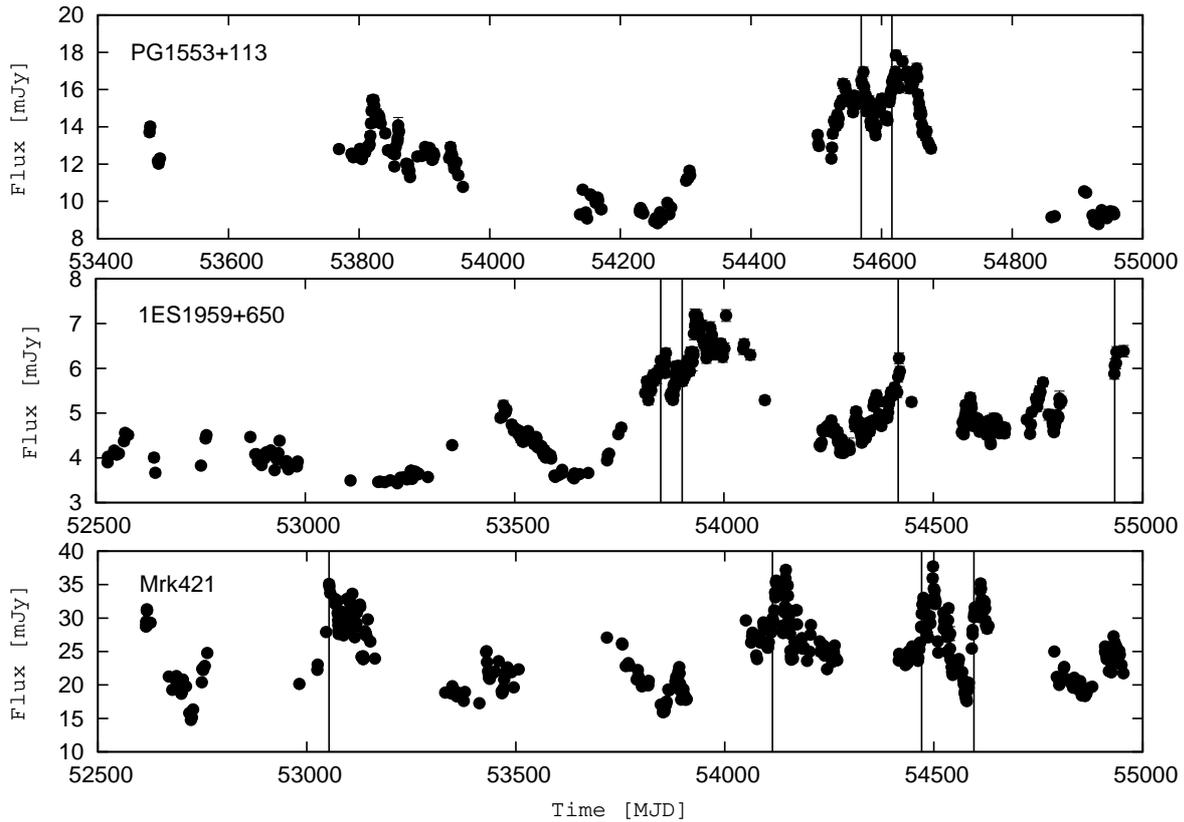}
\caption{Optical light curves of Mrk~421 (bottom), 1ES~1959+650 and PG~1553+113 from the Tuorla blazar monitoring program. The vertical lines in the light curves indicate the flares that fulfilled the triggering criteria of the MAGIC ToO proposal.}
\end{figure*}
 
\section{Tuorla Blazar Monitoring Program and the Analysis of the Optical Light Curves}

 The observations are performed with the KVA 35 cm telescope, which is
remotely operated from Tuorla Observatory and the Tuorla 1 meter
telescope. The observations are made in the R-band and the magnitudes are
measured using differential photometry with calibration stars in
the same CCD frame as the object. The light curves are updated to
the project web page http://users.utu.fi/kani/1m/ on a daily basis.

The long term monitoring program consists of 24 objects, which were
chosen from 
list of candidate TeV
blazars \cite{costamante} with declination $>20^o$ for them to be observable from
Tuorla. The monitoring started in 2002 and the goal of the monitoring
is to study the optical variability of these sources and to determine
the baseline flux. We have also made study of the host galaxy
contribution to the objects overall brightness \cite{nilsson}. 
This way we can get direct measurement of the AGN core
brightness by subtracting the host galaxy distribution. The collected
information has been used to define the trigger criteria for
Target of Opportunity program. As the program has been proven to be
successful (discovery of three new VHE $\gamma$-ray emitters following the optical alerts), we have added $\sim20$ more blazars to the monitoring
program. For analysis we chose 29 best sampled optical light curves (23
from the original sample and 6 additional ones: S5~0716+714, W~Com,
3C~279, PG~1553+113, H~1722+119, PKS~2155-304).

To define the flaring states from the optical light curves, we have
used the criteria of 50\% of increase of the core flux compared to
quiescent level. The quiescent flux level has been determined by
visual inspection of light curves, but has generally followed the rule
that source spends $\sim 20\%$ of the time below and within this
level. A more detailed study of the quiescent states and definition
of flaring states is in preparation, but following this simple
criterium our light curves reveal 53 optical flares.

\section{Optical to VHE $\gamma$-ray comparison}

The MAGIC telescope \cite{crab} is performing Target of Opportunity
observations of blazars when the blazars are in a high optical
state. Therefore, for quite many flares simultaneous or
quasi-simultaneous optical data exists. Of the total of 53 optical
flares, 43 occured after the beginning of Cycle 1 in MAGIC (June
2005). For 30 flares there is some simultaneous MAGIC data, but only
for 18 flares there is more than 3 hours of good quality data. This is
mostly constrained by the weather and moon conditions as well as
conflicts in the MAGIC schedule. In 11 cases the source was detected
by MAGIC. These 11 flares were detected in 9 sources: Mrk~421,
1ES~1959+650, PG~1553+113, 3C~66A, 3C~279, BL~Lac, Mrk~180,
1ES~1011+496, S5~0716+714 (see Fig.~2). In 7 cases the MAGIC
observations resulted with upper limit.

\begin{figure}[t]
\includegraphics[width=0.45\textwidth,angle=0,clip]{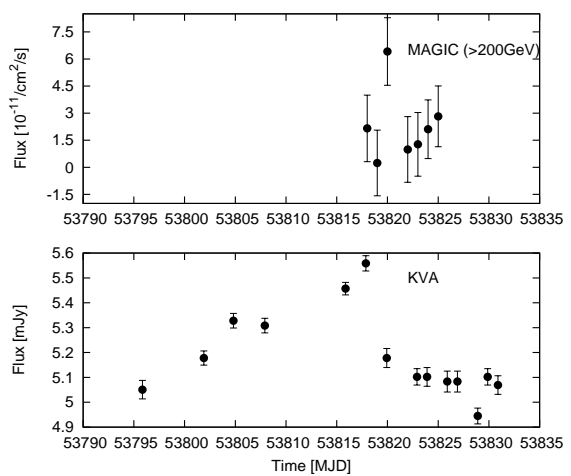}
\caption{Light curve of Mrk~180 from MJD 53790-53835 (2006, February 24 - April 10). {\it Lower panel}: the optical light curve from the KVA telescope. {\it Upper panel}: the MAGIC light curve.}
\end{figure} 

The only source for which a real correlation study is possible is 
Mrk~421, because it is the only one for which we have good time coverage 
with MAGIC. Preliminary results on this study will be presented 
\cite{monitoring}. For 1ES~1959+650 the optical flare occured in spring 2006 (MJD $\sim 53800$, see Fig.~2), 
when the source was in a very low state at VHE $\gamma$-ray regime \cite{tagliaferri}. 
PG~1553+113 was in high state in optical in spring 2008, when it was 
also detected to have a higher VHE $\gamma$-ray flux than in 2007 (in lower 
optical state) \cite{prandini}. We exclude 3C~66A 
from our comparison, it is discussed in \cite{errando}, but it should be 
mentioned that the discovery of VHE $\gamma$-rays from 3C~66A by VERITAS 
was done during a optical flare \cite{acciari}.

For other sources we need to compare single discovery observations and
the follow-up or previous observations in lower optical state. There
is now a second observation by MAGIC for all the discovered sources ,
which has been performed in a lower optical state. For Mrk~180 the
analysis of the follow-up observation from October-December 2008 is
still ongoing. For 1ES~1011+496 the observation from 2006 \cite{hbl}
showed only weak evidence for a VHE $\gamma$-ray signal at
3.5$\sigma$. The VHE $\gamma$-ray flux was $>40\%$ higher in March-May
2007 than in March-April 2006. For S5~0716+714 the November 2007 flux
(low optical state) is about 8 times lower than April 2008 flux (high
optical state) \cite{0716}. BL~Lac was only detected in 2005
observation period but not in 2006, when it was in signicantly lower
optical state. The case for 3C~279 is more complicated. In February
2006 when MAGIC first discovered it \cite{3C279}, it was in a rather
high optical state, but not fulfilling the triggering criteria.  In
January 2007, when the observations were triggered by the optical high
state (and the optical flux was three times higher that in February
2006), a preliminary analysis shows a detection the source by MAGIC \cite{berger}. As the analysis
of the 2007 data is still ongoing, the comparison of flux levels
between February 2006 and January 2007 is not yet possible.

\begin{figure}[!h]
\includegraphics[width=0.25\textwidth,angle=270,clip]{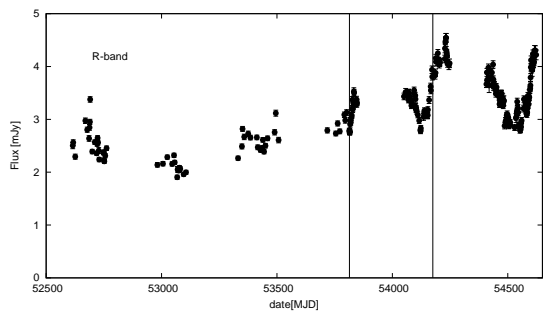}
\caption 
{The optical light curve of 1ES~1011+496 shows six years (2003-2008) of data from the Tuorla blazar monitoring program. The vertical lines indicate the starting points of the MAGIC observations in 2006 \cite{hbl} and 2007 \cite{1011}}. 

\end{figure}

\begin{figure}[!h]
\includegraphics[width=0.45\textwidth,angle=0,clip]{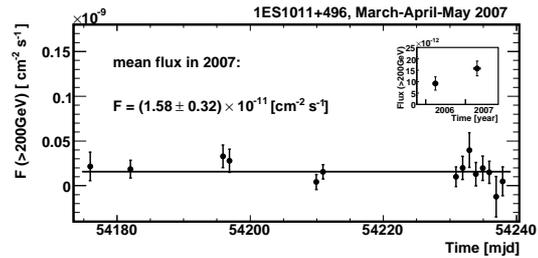}
\caption
{The VHE $\gamma$-ray light curve of 1ES~1011+496 from March 17th (MJD 54176) to May 18th (54238) 2007. 
The inset shows the yearly averages from 2006 and 2007.}
\end{figure}

For non-detected sources the results from the MAGIC observations have 
not been published yet, with the exception of OJ~287 during the December 
2007 outburst \cite{seta,oj287}. However, for non-detected sources 
even the flaring state VHE $\gamma$-ray flux might be below the 
MAGIC sensitivity. Therefore, the only source which shows clear 
evidence that the VHE $\gamma$-ray flux does NOT follow the optical 
flux states is 1ES~1959+650. Interestingly, for this source also lack of correlation between X-rays and VHE $\gamma$-rays has been observed \cite{kraw}. 
 
\section{Discussion and Outlook}

The optically triggered Target of Opportunity observations have
resulted in the discovery of three new VHE $\gamma$-ray emitting
blazars. Mrk~180 and 1ES~1011+496 are classified to be HBLs, while
S5~0716+714 is classified as IBL or even LBL. However, the BL Lac
population is continuous rather than bimodal or trimodal (see,
e.g. \cite{padovani,nieppola}), and in all of the three discovered sources the
synchrotron peak frequency is not too far from the optical band (see
e.g. publicly available data at {\it asdc.asi.it/blazars} and
Fig.~6. in \cite{1011}). If the synchrotron emission and the IC
emission arise from the same emission region, the connection between
optical and VHE $\gamma$-rays would be expected. On the other hand, it 
can be that in some sources there is no connection between optical and 
VHE $\gamma$-ray flaring (e.g. different emission 
regions, different mechanisms etc.).

There seems to be at least 6 sources which suggest connection between 
optical and VHE $\gamma$-ray high states and only one 
source where clearly no connection is seen. Unfortunately the current 
data from VHE $\gamma$-ray range does not allow more definite 
conclusions.

\section*{Acknowledgments}
  MAGIC enjoys the excellent working conditions at ORM and is
  supported by the German BMBF and MPG, the Italian INFN, the Spanish
  MCI-NN, ETH research grant TH 34/04 3, the Polish MniSzW Grant N
  N203 390834, and by the YIP of the Helmholtz Gemeinschaft. E.J.L
  wishes to acknowledge the support by the Academy of Finland grant
  127740. D.M's research was supported by a Marie Curie
  Intra European Fellowship within the 7th European Community
  Framework Programme.

\end{document}